\documentclass[a4paper]{article}
\usepackage{graphicx}
\usepackage{amssymb}
\usepackage{latexsym}
\usepackage{cite}
\usepackage{a4wide}

\newcommand{\pbar}{\mbox{$\overline{\mathrm p}$}}
\newcommand{\Hbar}{\mbox{$\overline{\mathrm H}$}}

\newcommand{\pbhe}{$\overline{\mathrm p} {\rm He}^{+}$}

\newcommand{\pbhef}{$\overline{\mathrm p} ^4{\rm He}^{+}$}
\newcommand{\pbhet}{$\overline{\mathrm p} ^3{\rm He}^{+}$}
\newcommand{\Se}{$\vec{S}_e$}
\newcommand{\Lp}{$\vec{L}_{\overline{\mathrm p}}$}
\newcommand{\Sp}{$\vec{S}_{\overline{\mathrm p}}$}

\newcommand{\Lps}{$L_{\overline{\mathrm p}}$}
\newcommand{\mup}{$\vec{\mu}_{\overline{\mathrm p}}$}

\newcommand{\nuHFp}{$\nu_{\mathrm{HF}}^+$}
\newcommand{\nuHFm}{$\nu_{\mathrm{HF}}^-$}

\newcommand{\nuSHFp}{$\nu_{\mathrm{SHF}}^+$}
\newcommand{\nuSHFm}{$\nu_{\mathrm{SHF}}^-$}

\begin{document}

\title{Precision Physics with Low-energy Antiprotons --
from AD to FLAIR%
\thanks{Presented at the XXXI Mazurian Lakes Conference, Piaski, Aug. 30 - Sep. 6, 2009. }%
}
\author{Eberhard Widmann\\
{\it Stefan Meyer Institute, Austrian Academy of Sciences},\\
{\it  Boltzmanngasse 3, 1090 Vienna, Austria}
}
\maketitle
\begin{abstract}
Experiments with low-energy antiprotons are currently performed at the Antiproton Decelerator of CERN. The main experiments deal with the spectroscopy of antiprotonic helium, an exotic three-body system, and the formation and spectroscopy of antihydrogen. A next generation facility FLAIR (Facility for Low-energy Antiproton Rsearch) is planned at the FAIR facility, generating a factor 100 higher flux of stopped antiprotons and also offering continuous antiprotons beam, which will enable nuclear and particle physics type experiments.  
\end{abstract}

\noindent 36.10.Gv,11.30.Er,14.20.Dh,39.30.+w
  
\section{Introduction}
The physics with low-energy antiprotons currently concentrates on precision spectroscopy of antiprotonic atoms and antihydrogen. The reason for that is given by the characteristics of the only low-energy antiproton beam available currently at the Antiproton decelerator (AD) of CERN. The AD \cite{Eriksson-LEAP03,Eriksson:2009fj} produces pulses of 3--5$\times10^{7}$ antiprotons of 5.3 MeV kinetic energy every 90--120 second, which makes the beam only usable to be trapped in Penning traps or stopped in low-density gas targets. Two collaborations working at the AD, ATRAP \cite{ATRAP:09} and ALPHA \cite{ALPHA:09}, have as goal to produce antihydrogen, the most simplest atom consisting only of antimatter, from its charged constituents by trapping antiprotons and positrons in Pennning traps, to trap the resulting antihydrogen in a neutral-atom trap and to perform 1S-2S laser spectroscopy. This transition is known in hydrogen to a precision of $\sim 10^{-14}$ \cite{Niering:00}, and a measurement in antihydrogen thus offers one of the best tests of CPT symmetry. The ASACUSA collaboration \cite{ASACUSA} proposed to measure the second best known quantity in hydrogen, the ground-state hyperfine splitting ($\Delta f/f\sim 10^{-12}$ \cite{Karshenboim:02}), in an atomic beam setup with \Hbar\ \cite{HbarLOI}.

In addition to the planned antihydrogen experiment, ASACUSA is pursuing the precision laser and microwave spectroscopy of antiprotonic helium, an exotic atomic system containing an antiproton. The experiment has led (under the assumption of CPT non-conservation) to the most precise determination of the antiproton mass and charge as well as its magnetic moment.

To overcome the limitations of the AD both in intensity and in the lack of fast extraction, {\em i. e.} continuous beam, a new facility FLAIR is planned at the FAIR facility in Darmstadt, Germany. Here both the currently ongoing experiments at the AD as well as new experiments could be performed.

This talk gives an overview on the current an planned activities with low-energy antiprotons at the AD and FLAIR.

\section{Precision spectroscopy of antiprotonic helium}

\begin{figure}[h]
\begin{center}
\includegraphics[scale=0.4]{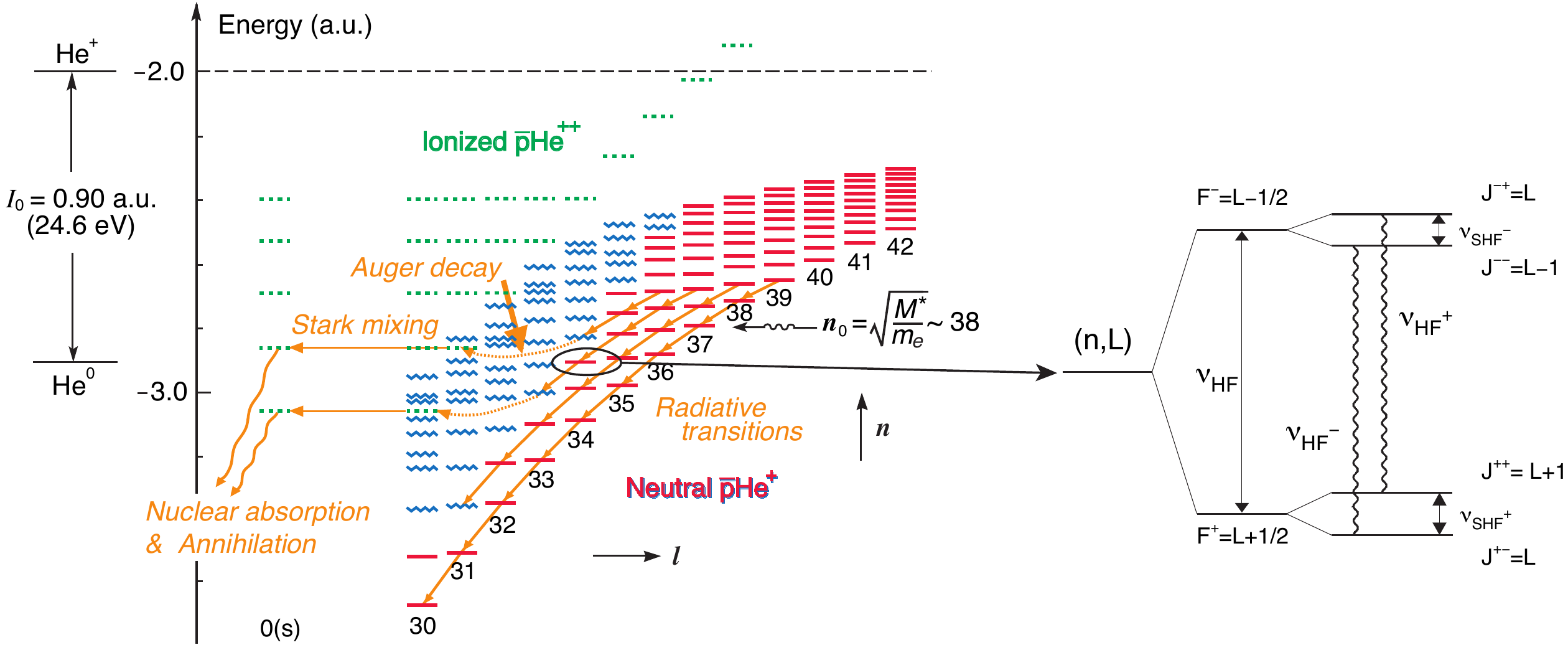}
\caption{Level diagram of antiprotonic helium. The solid lines correspond to meatastable states, the wavy ones to short-lived ones. Each of the levels in the left hand side is, in the case of antiprotonic $^4$He, split into a quadruplet due to the magnetic interaction of its constituents.}
\label{fig:pbhe}
\end{center}
\end{figure}

Antiprotonic helium (\pbhe) is an exotic three-body system consisting of an antiproton, a helium nuceleus, and an electron. It comprises a series of metastable states with microsecond life times (cf. Fig.~\ref{fig:pbhe}). Initially discovered at KEK in 1991 \cite{Iwasaki:91}, it has been studied extensively at LEAR \cite{Yamazaki:02} and later at the AD  \cite{Hayano:2007}.

\subsection{Laser spectroscopy and the antiproton mass}

\begin{figure}[h]
\begin{center}
\begin{minipage}{0.45\textwidth}
\includegraphics[scale=0.13]{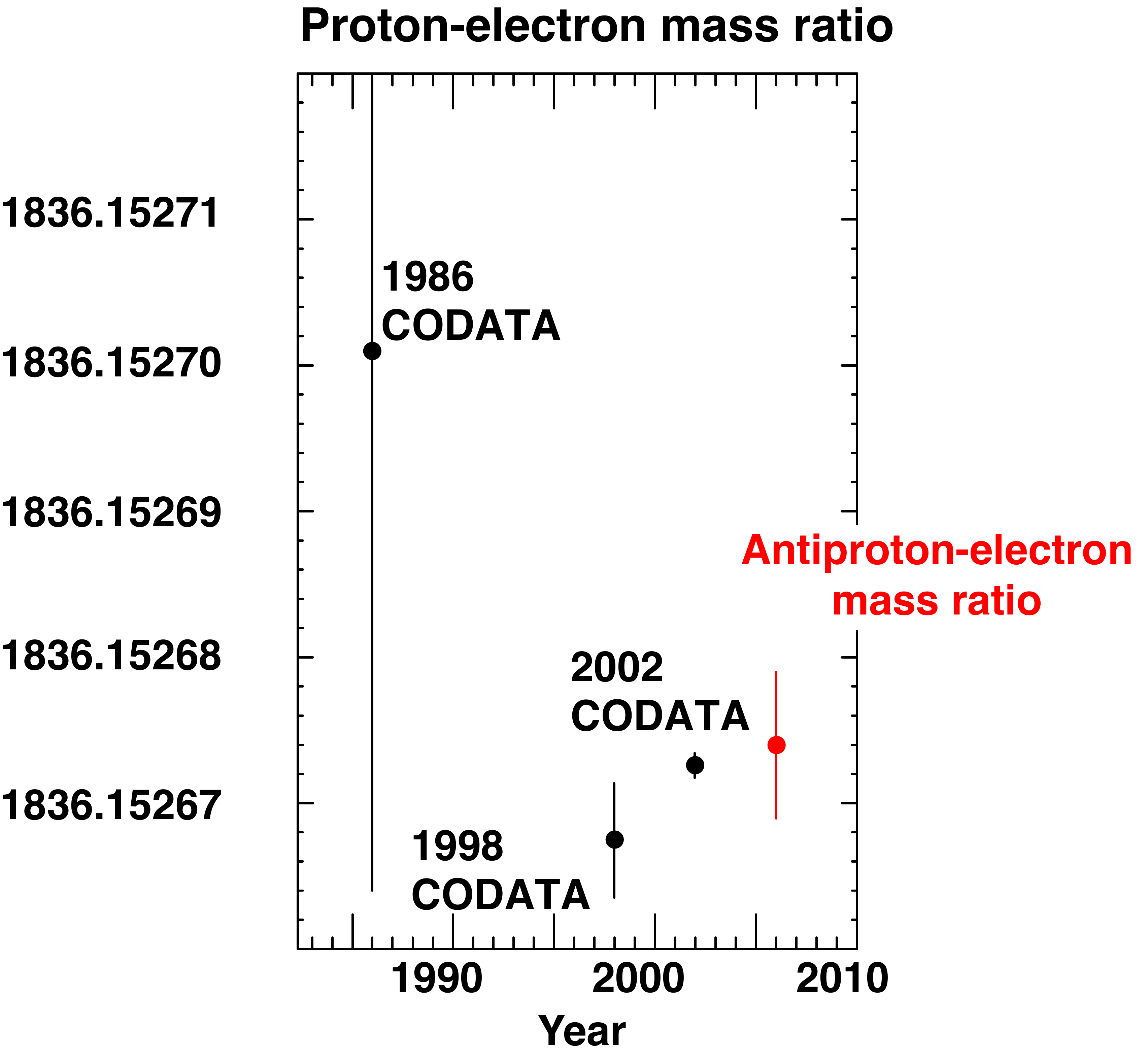}
\end{minipage}  
\begin{minipage}{0.45\textwidth}
\includegraphics[scale=0.2]{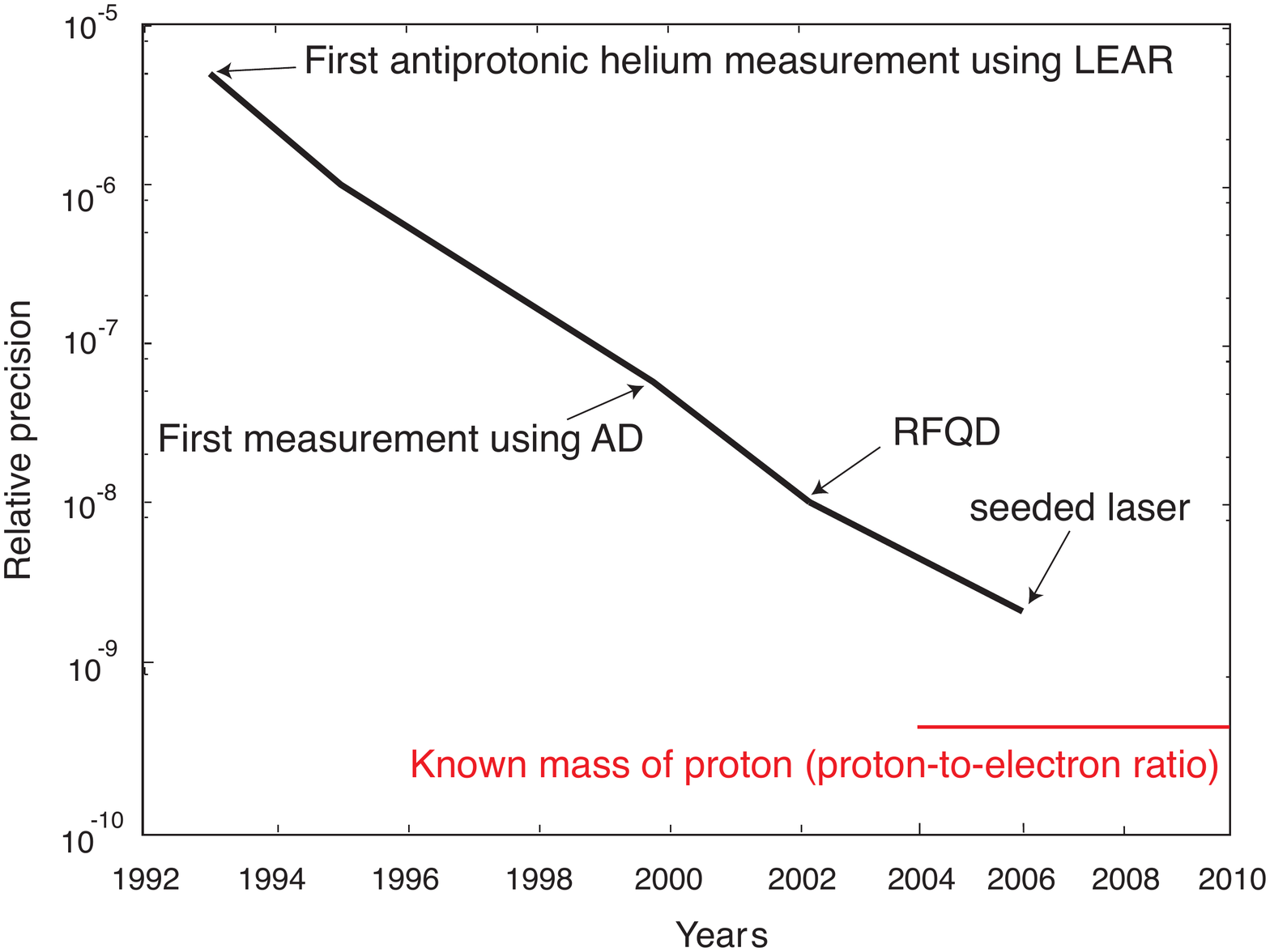}
\end{minipage}
\caption{Left: Antiproton-to-electron mass ratio as obtained by ASACUSA compared to the equivalent value of the antiproton. Right: Relative precision of the CPT test of the antiproton mass as a function of time.}
\label{fig:pbarmass}
\end{center}
\end{figure}

Using lasers, transitions of the antiproton can be excited between metastable and short-lived states, providing an easy detection of a resonance condition by the immediate annihilation on the antiproton if it reaches a short-lived state. Making use of this technique, a total of 13 transitions in \pbhef\ and \pbhet\ could be measured with increasing precision \cite{Hori:01,Hori:03,Hori:2006}. As shown in Fig.~\ref{fig:pbarmass} right, one major increase of precision was possible by using an Radio Frequency Quadrupole Decelerator (RFQD) \cite{RFQD-01} to decelerate antiprotons from 5.3 MeV to about 100 keV, making it possible to stop antiprotons in a dilute gas and thus completely getting rid of systematic errors due to pressure shifts. The last improvement came through the development of a new pulse-amplified cw laser system \cite{Hori:2006}.

By comparing the laser spectroscopy data to precise three-body QED calculations \cite{korobovexa05,Kino:04}, limits on the relative difference of the proton and antiproton mass $M$ and charge $Q$ can be obtained. The currently most accurate value \cite{Hori:2006} is 
\begin{equation}
\frac{Q_{\overline{\mathrm p}}+Q_{\mathrm p}}{Q_{\mathrm p}}
\sim \frac{M_{\mathrm p}-M_{\overline{\mathrm p}}}{M_{\mathrm p}}
\leq 3 \times 10^{-9},
\end{equation}
which is one of the most precise tests of CPT symmetry in the baryon sector.

\subsection{Microwave spectroscopy and the antiproton magnetic moment}

The hyperfine structure (HFS) of \pbhe\ as depicted in Fig.~\ref{fig:pbhe} can be used to determine the magnetic moment \mup\ of the antiproton. The HFS is unique because it is generated by three magnetic moments: the ones related to the electron spin \Se, the {\em orbital} angular moment of the antiproton \Lp, and the \pbar\ spin \Sp. Because of the large angular momentum of the \pbar\ in metastable states (\Lps\ $\sim 35$), the largest splitting is caused by the interaction of the electron spin with the antiproton angular momentum, \Lp $\cdot$\Se, causing a {\em hyperfine} structure. The interaction of the antiproton spin magnetic moment \mup\ with the other moments causes a second, about a factor 100 smaller splitting called {\em super-hyperfine} splitting.

\begin{figure}[h]
\begin{center}
\begin{minipage}{0.45\textwidth}
\includegraphics[scale=0.4]{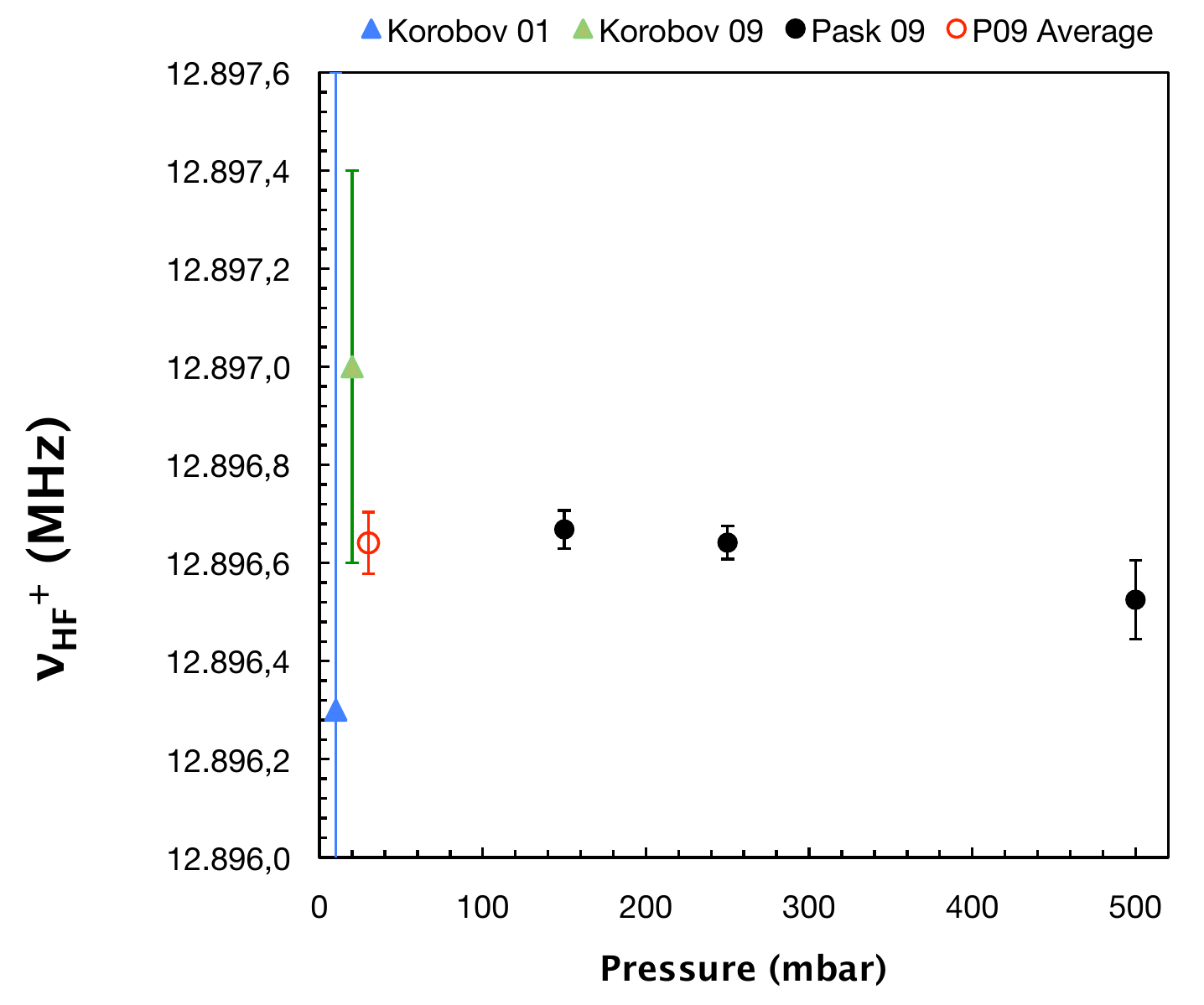}
\end{minipage} \hspace*{2mm}
\begin{minipage}{0.45\textwidth}
\includegraphics[scale=0.4]{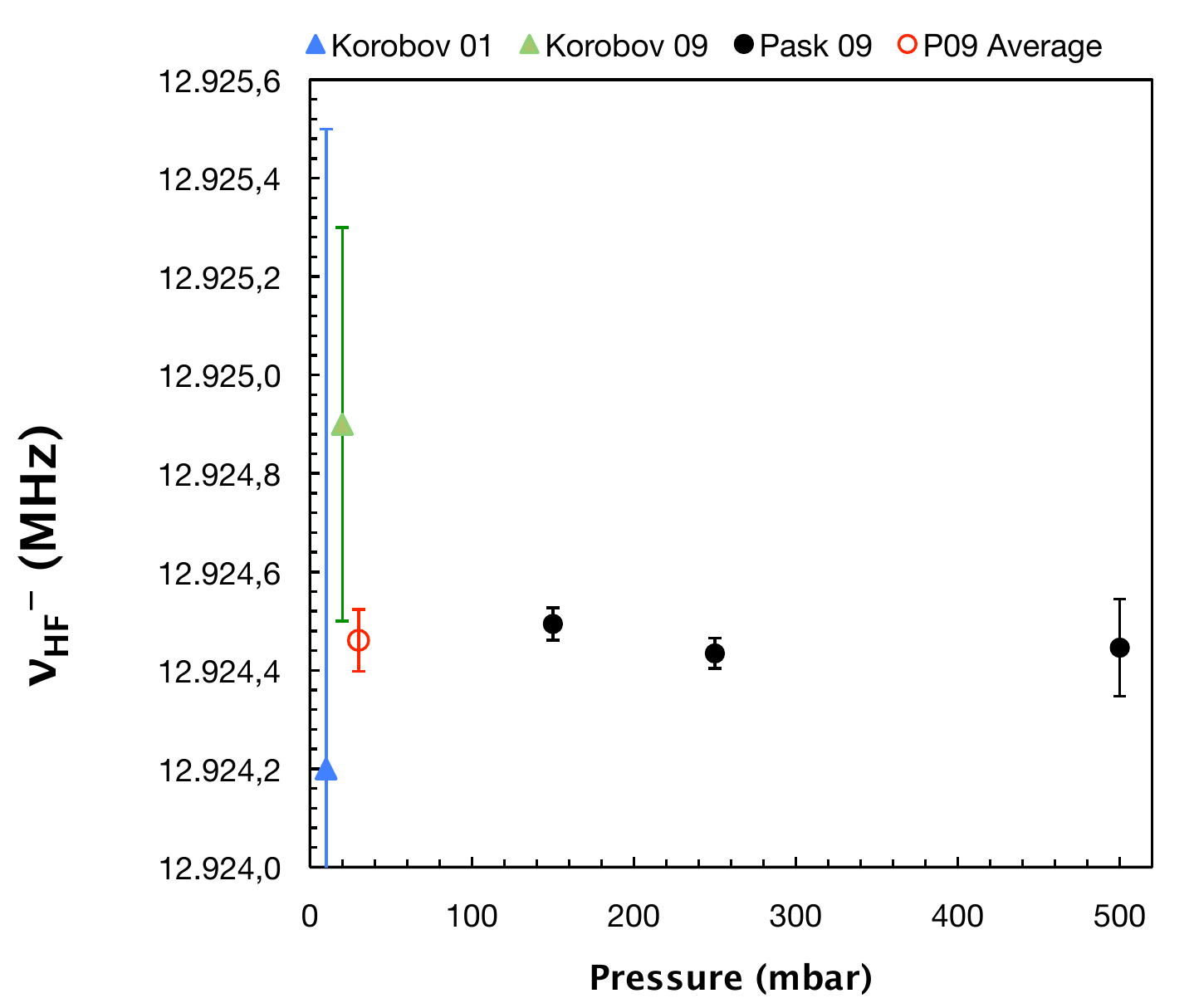}
\end{minipage}
\caption{Comparison of measurement and theory for the hyperfine transitions \nuHFp\ (left) and \nuHFm\ (right) for the $(37,35)$ state of \pbhef. The full circles are data points taken at different densities, the open circle shows the density-averaged experimental value (Pask 09: \cite{Pask:2009lq}). The open and closed triangles represent theoretical values (Korobov 01: \cite{Korobov:01}, Korobov 09: \cite{Korobov:2009dq}). The errors of theory contain both the numerical uncertainty and estimated higher order corrections.}
\label{fig:nuHF}
\end{center}
\end{figure}

Within ASACUSA we devised a laser-microwave-laser resonance method \cite{Widmann:02} to measure the two transitions \nuHFp\ and \nuHFm\ depicted by curly lines in Fig.~\ref{fig:pbhe}. The two transitions are caused by an electron spin flip and are therefore not directly sensitive to \mup\ \cite{Bakalov:07}. Fig.~\ref{fig:nuHF} gives the comparison of measured values  and theoretical predictions, showing agreement within the error bars with both calculations. The experimental values (typical error: $5\times10^{-6}$) are significantly more accurate than the theoretical ones ($3\times10^{-5}$ \cite{Korobov:2009dq}).

\begin{figure}[h]
\begin{center}
\includegraphics[scale=0.3]{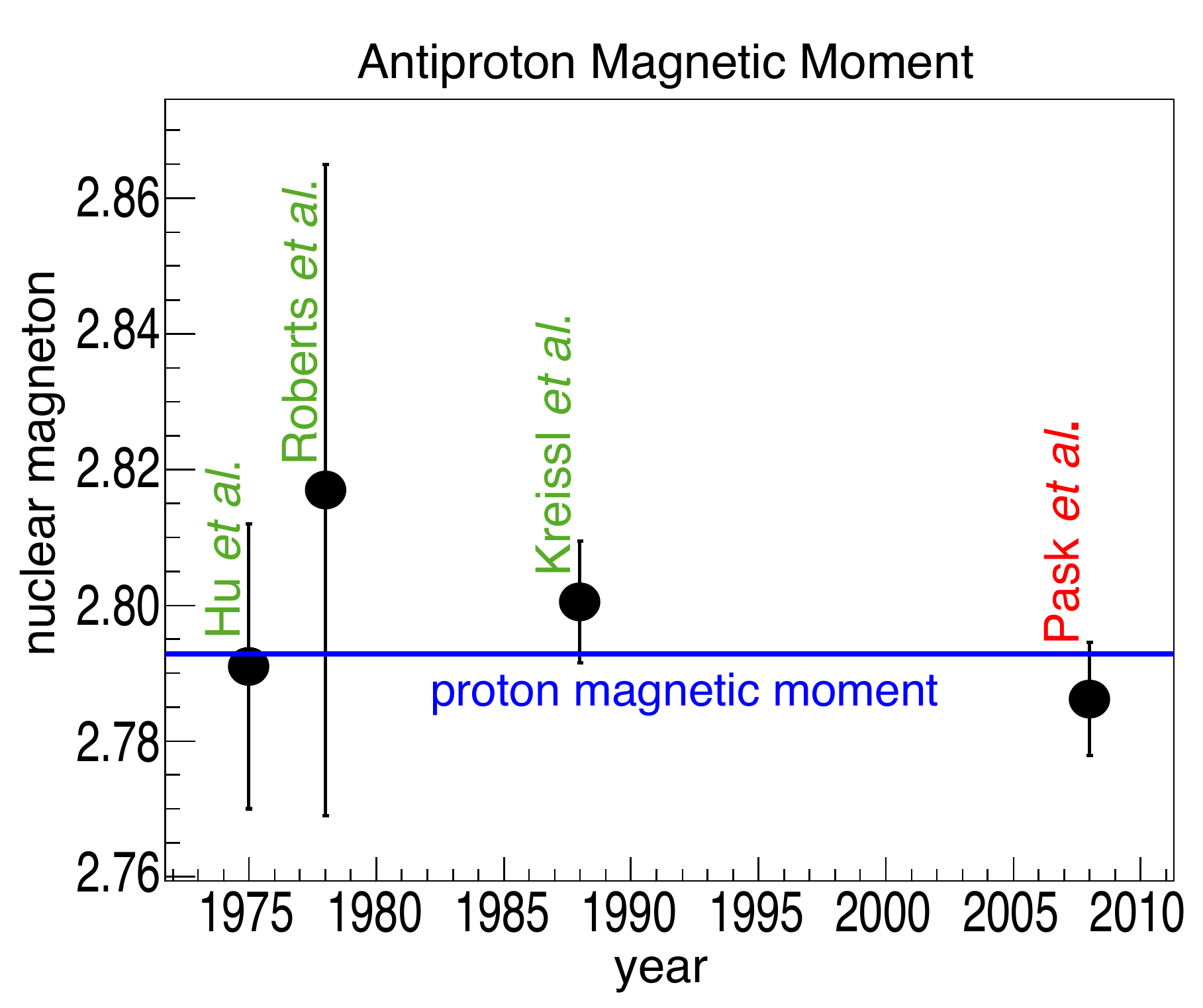}
\caption{Determination of the magnetic moment of the antiproton as a function of time. Hu {\em et al.}: \cite{Hu:75}, Roberts {\em et al.}: \cite{Roberts:75}, Kreissl {\em et al.}: \cite{Kreissl:88}, Pask {\em et al.}: \cite{Pask:2009lq}.}
\label{fig:mupbar-th-exp}
\end{center}
\end{figure}

In order to learn about the magnetic moment of the antiproton, the most sensitive quantity is the difference $\Delta \nu$ = \nuHFm\ -- \nuHFp\ = \nuSHFp\ -- \nuSHFm. Since the two lines \nuHFp\ and \nuHFm\ are very close together ($\Delta \nu \sim$ 28 MHz for $(n,L)=(37,35)$), the experimental error on the difference is rather large. The final result is \cite{Pask:2009lq}
\begin{equation}
\frac{\mu_{s}^{p}-|\mu_{s}^{\overline{p}|}}{\mu_{s}^{p}} = (2.4\pm2.9)\times 10^{-3},
\end{equation}
slightly better than the previous value of PDG \cite{PDG:08}. Fig.~\ref{fig:mupbar-th-exp} provides a comparison of our new value with older ones, showing good agreement with the value for the proton.

\section{Antihydrogen}

\begin{figure}[h]
\begin{center}
\includegraphics[scale=0.4]{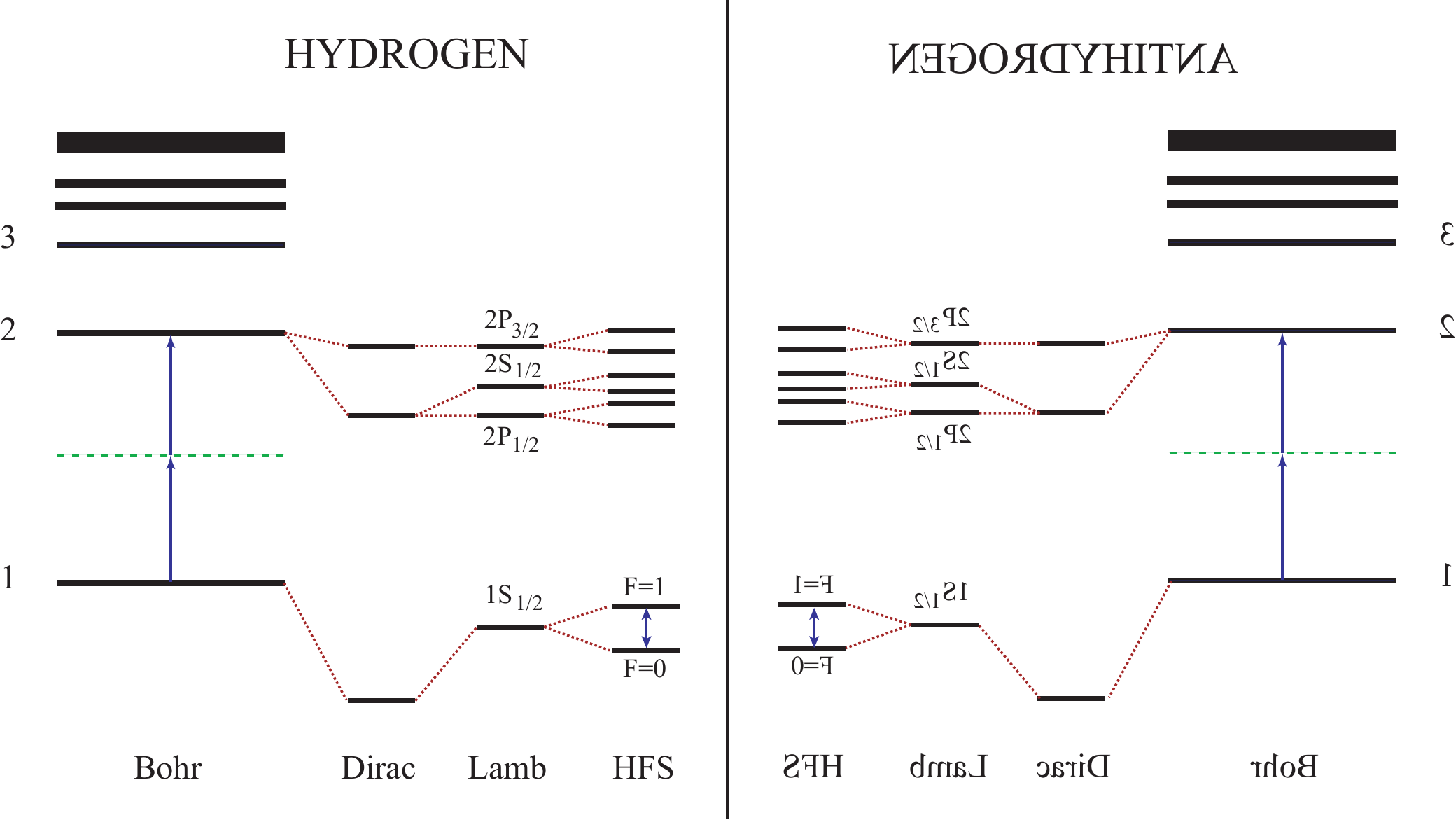}
\caption{Level diagram of hydrogen and antihydrogen.}
\label{fig:hbar}
\end{center}
\end{figure}

The fascination of antihydrogen lies in the fact that it is the simplest form of pure antimatter (a bound state of an antiproton and a positron). It was the big media echo after the discovery of the first 9 antihydrogen atoms at LEAR in 1996 \cite{Baur:96} that made the construction of the AD possible. CPT symmetry predicts that all properties including the optical spectra of antihydrogen are identical to those of hydrogen (cf. Fig.~\ref{fig:hbar}). 

\begin{figure}[h]
\begin{center}
\includegraphics[scale=0.4]{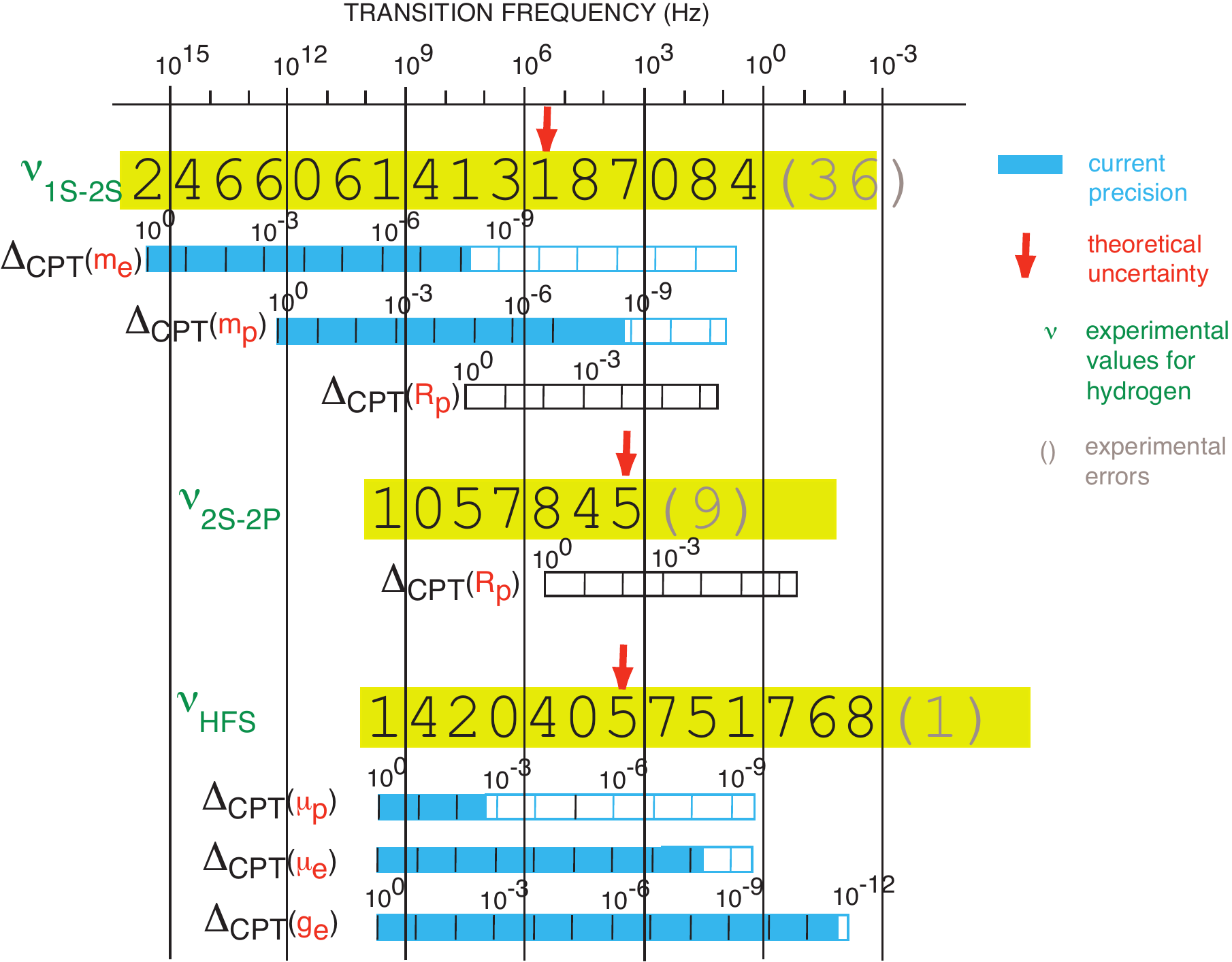}
\caption{Measured quantities of hydrogen and their relation to CPT variables. Experimental values: $\nu_{\mathrm{1S-2S}}$: \cite{Niering:00},  $\nu_{\mathrm{2S-2P}}$: \cite{De-Beauvoir:2000ru}, $\nu_{\mathrm{HFS}}$: \cite{Karshenboim:02}. The relative accuracies of properties of the positron and antiproton are taken from PDG \cite{PDG:08}. }
\label{fig:hbar-comp}
\end{center}
\end{figure}

Among the best-known quantities of hydrogen are the 1S-2S two-photon transition and the ground-state hyperfine structure (GS-HFS), both known to very high precision. A direct effect of QED is the Lamb shift, but its use for high-precision comparison is limited due to the large level width caused by the short life time of the 2P state. Fig.~\ref{fig:hbar-comp} shows the three quantities on an absolute scale, with their experimental errors and their sensitivities to properties of the antiproton and the positron. Obviously $\nu_{\mathrm{1S-2S}}$ and $\nu_{\mathrm{HFS}}$ test different interactions (1S-2S: electric, HFS: magnetic interactions) and have a very different impact on the knowledge of the CPT properties of the constituents. Also the precision of QED theory is shown, which is limited by a hadronic correction: the experimentally not well known proton radius. If the experimental accuracy exceeds the limit of theory, also information on the equality of the internal structure of proton and antiproton can be obtained.

\begin{figure}[h]
\begin{center}
\begin{minipage}{0.45\textwidth}
\includegraphics[scale=0.35]{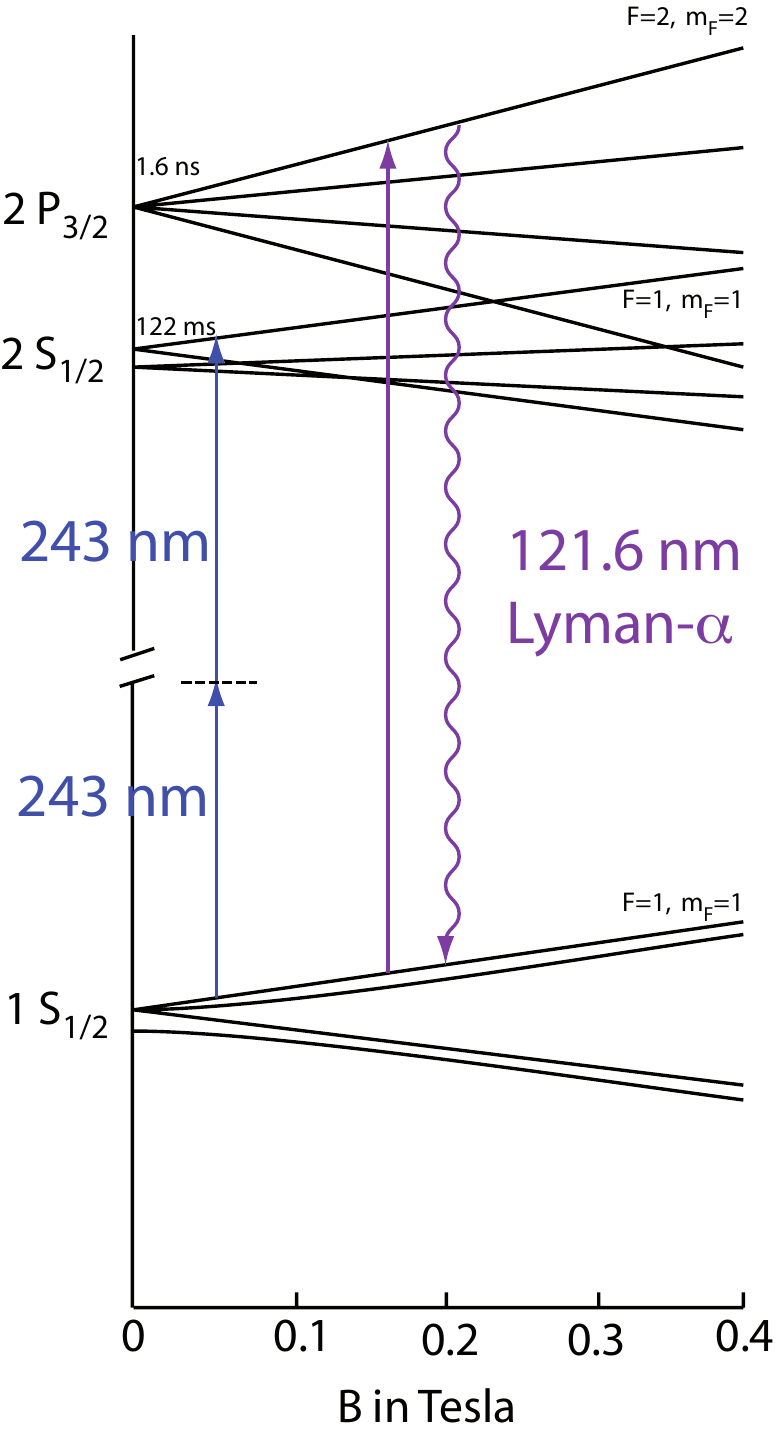}
\end{minipage}
\begin{minipage}{0.45\textwidth}
\includegraphics[scale=0.45]{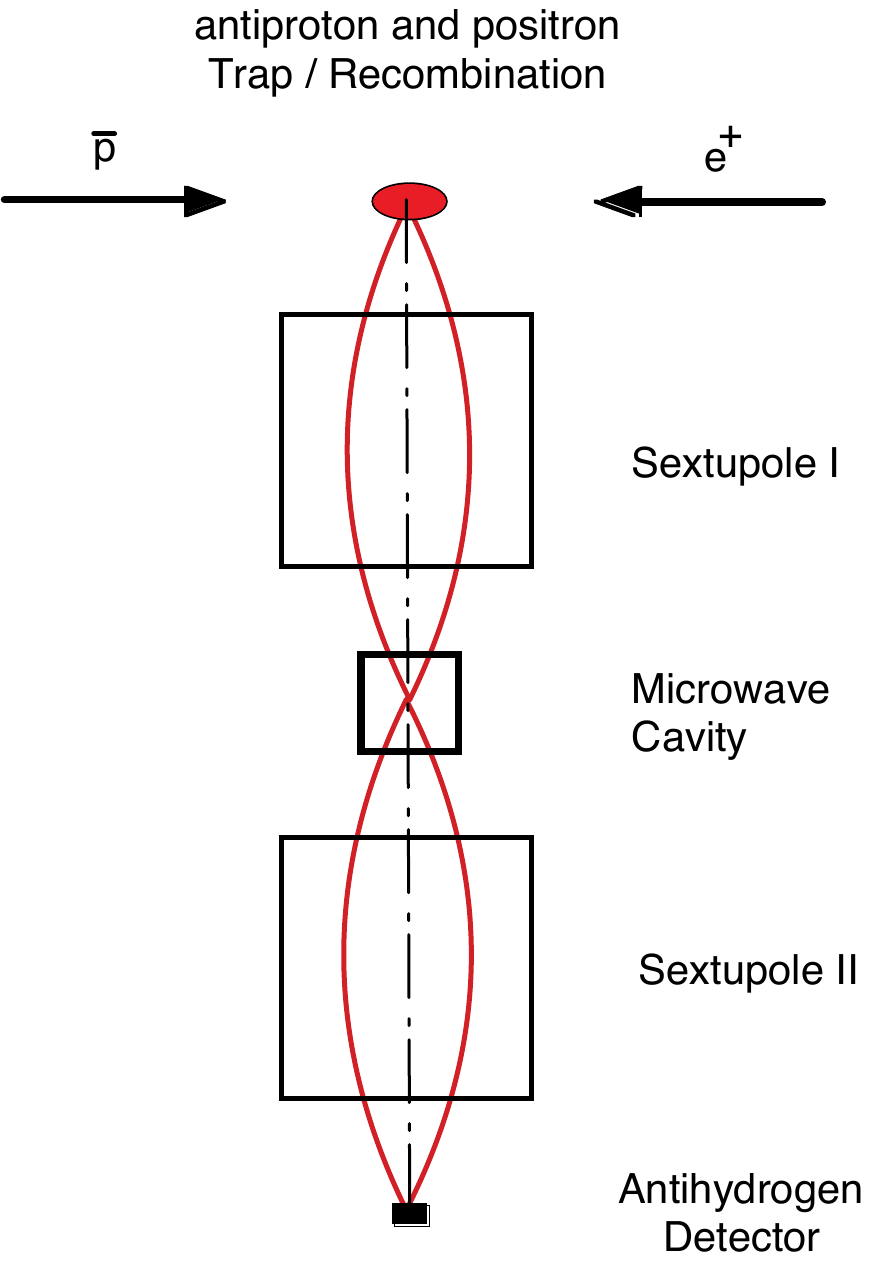}
\end{minipage}
\caption{Proposed measurements with antihydrogen: 1S-2S two-photon laser spectroscopy (left) and an atomic beam line to measure the ground-state hyperfine structure (right).}
\label{fig:hbaroverview}
\end{center}
\end{figure}

The two collaborations ATRAP and ALPHA (the successor of ATHENA), are pursuing the goal to measure $\nu_{\mathrm{1S-2S}}$ of antihydrogen to highest precision (Fig.~\ref{fig:hbaroverview} left). Both published the first production of cold antihdrogen in 2002 \cite{ATHENA-Hbar:02,ATRAP-Hbar:02a} employing the same method of nested Penning traps described in more detail in \cite{Gabrielse:2005sl}. The goal here is to make antihydrogen cool enough to be trapped in neutral atom trap \cite{Gabrielse:2005sl}. Both collaborations have built Ioffe-Pritchard type neutral atom traps on top of the Penning traps needed for the charged constituents of antihdrogen and succeeded in forming antihydrogen in such configurations \cite{Gabrielse:2007ai,Gabrielse:2008nx,Andresen:2008wh}, but no trapping of neutral antihydrogen atoms has been reported so far, showing the long-term nature of such fundamental physics projects. 

A new experiment AEgIS \cite{AEGIS} has been approved recently that aims at a first measurement ever of the gravitational acceleration of antimatter. Within ASACUSA, we are planning to measure $\nu_{\mathrm{HFS}}$ of antihydrogen using an atomic beam method \cite{HbarLOI} (cf. Fig.~\ref{fig:hbaroverview} right) as was done in the early days of hydrogen. This method has the advantage, that no trapping is needed and that antihydrogen atoms with temperatures of 100 K can be used. Simulation show  that with about 100 \Hbar /s in the ground state, a relative precision of $10^{-7}$ can be reached \cite{Juhasz:ve}.

In order to provide \Hbar\ suitible for the atomic beam line, ASACUSA develops two new methods of antihydrogen formation: 
\begin{enumerate}
\item  a {\em cusp} trap \cite{MohriEP}, which is a magnetic bottle like structure known to also be able to trap neutral atoms. The cusp trap is currently running at AD and has achieved trapping of antiprotons and protons. Antihydrogen formation has been attempted for the first time this year, but no result is available yet. Due to the strongly inhomogenous magnetic field of the cusp trap, the antihydrogen atoms will be polarized when leaving the formation region and therefore the first sextupole can be emitted (cf. Fig.~\ref{fig:hbaroverview} right). A superconducting sextupole will be delivered early next year and a spin-flip cavity is currently under design at CERN \cite{Juhasz:ve}, so that first experiments can start even next year.

\item A {\em Paul} trap \cite{HbarLOI} is under development at CERN and MPQ Munich. This trap has the advantage that it can produce \Hbar\ in a very small volume of $\sim 1$ mm$^{3}$, thus being a point source for very efficient transport through the beam line. 
\end{enumerate}

\section{The FLAIR facility}

\begin{figure}[h]
\begin{center}
\includegraphics[scale=0.6]{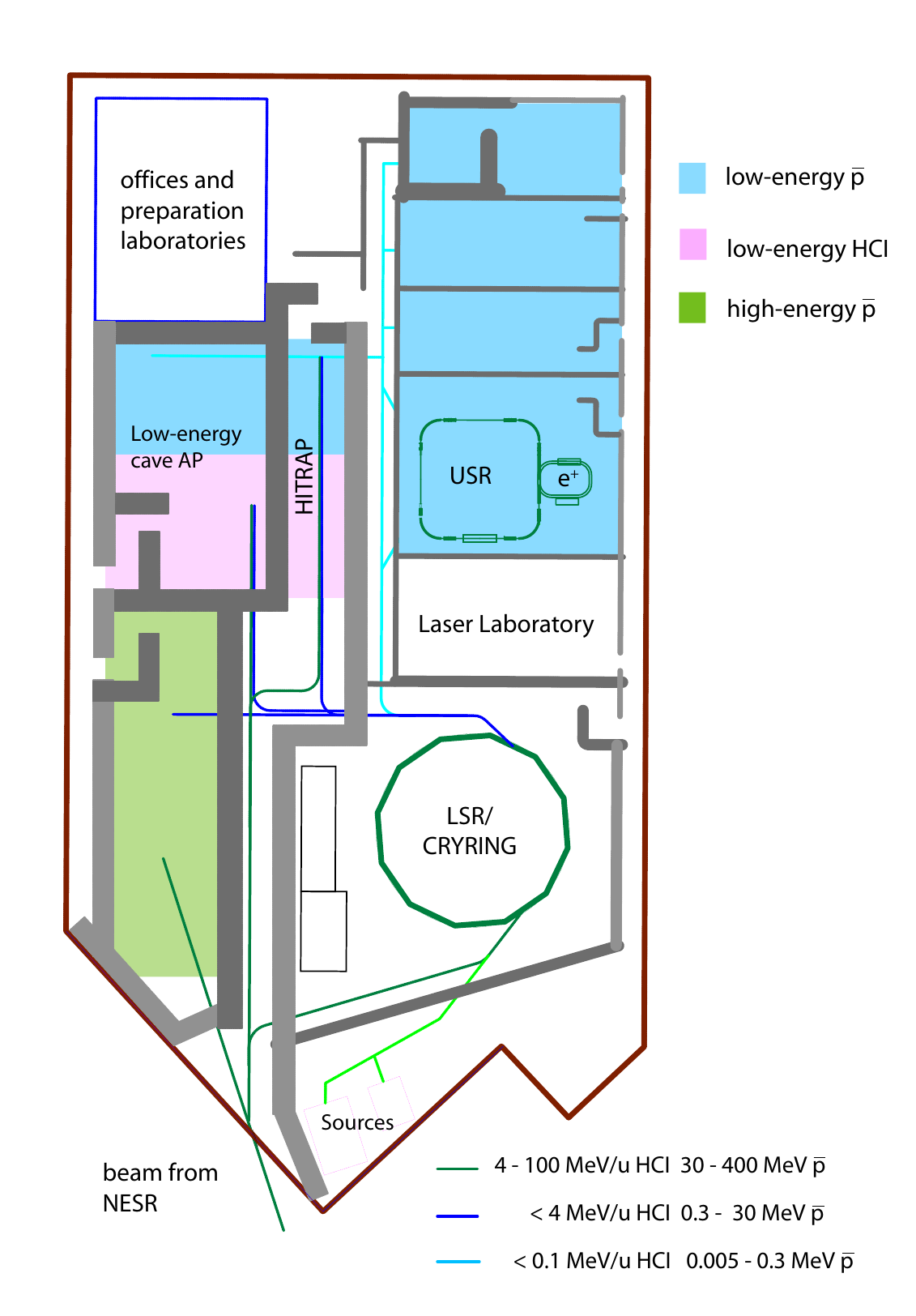}
\caption{Layout of the proposed FLAIR facility}
\label{fig:FLAIR}
\end{center}
\end{figure}

To overcome the limitations of the AD at CERN a next generation low-energy antiproton facility was proposed with the name of FLAIR \cite{FLAIR} -- Facility for Low-energy Antiproton and Ion Research -- for the planned FAIR facility \cite{FAIR} at Darmstadt, Germany. As spelled out in the FLAIR Letter of Intent \cite{FLAIR-LOI}, the new facility should have antiproton beams at 100 times lower energy than the AD using two storage rings, the {\em LSR} going to 300 keV and the {\em USR}, going to 20 keV (cf. Fig.~\ref{fig:FLAIR}). Both rings should have electron cooling and provide both slow and fast extracted beams. For LSR, the FLAIR collaboration chose the existing CRYRING \cite{Danared:2009bz} of Manne Siegbahn Laboratory, Stockholm, Sweden, which fulfills all requirements and will be contributed by the Swedish government as an in-kind contribution to FAIR. The USR \cite{Welsch:2009nk} is an electrostatic storage ring that is currently under development at GSI, Germany and Cockcroft Institute, UK.

The physics program of FLAIR covers currently the following topics \cite{Widmann:2005ys}
\begin{itemize}
\item Precision spectroscopy for tests of CPT and QED.\\
The experiments with antihydrogen, judging from the current speed of progress and the typical time scales for hiph precision fundamental experiments, will take more than one decade to complete. FLAIR would surely speed up the advance by providing a higher rate of stopped antiprotons. Especially experiments using laser-cooled antiprotons for spectroscopy and for gravitational measurements \cite{grav:2003} will take a long time to accomplish.
\item Atomic collisions.\\
The USR with an internal target and a reaction microscope is an ideal tool to study atomic collision processes at extremely low energies. Antiproton beams offer shorter pulses that any laser can deliver at the moment to study sub-femtosecond correlated dynamics in atoms. 
\item Antiprotons as hadronic probes.\\
The slow extracted beams of FLAIR would make nuclear and particle physics type experiments possible. This includes the X-rays spectroscopy of light antiprotonic atoms to study the nucleon-antinucleon interactions as well as X-rays measurements of heavy nuclei to study neutron halo effects. An especially interesting possibility will be the extension of these measurements to unstable nuclei which is proposed in the Exo+pbar experiment. A new idea is to study double-strangeness production in antiproton annihilation \cite{Zmeskal:2009ix} to verify data of OBELIX indicating a unexpectedly high production rate at LEAR and to investigate the possible formation of anikaon nuclear bound states with two $K^{-}$.
\item Medical applications.\\
In addition to protons and carbon ions currently used for cancer therapy with ion beams, antiprotons offer the additional deposit of energy released from the annihilation process. An experiment at CERN-AD, ACE, is currently investigating this possibility \cite{Fahimian:2009zt}. 
\end{itemize}

\section{Summary}

Currently low-energy antiprotons provide some of the most precise tests of CPT symmetry and three-body QED calculations. Antihydrogen experiments are likely to provide the best tests of CPT symmetry and first measurements of the gravitational interaction of antimatter. FLAIR would improve the conditions for theses experiments and allow to perform others, especially hadron physics related ones, not possible at the moment. The recent developments of FAIR have led to a constellation where FLAIR is not in the start phase but would be added at a later stage. For the low-energy physics community it is essential that the AD at CERN will remain operational in the mean time. Efforts are under way to install an additional ring called ELENA at the AD \cite{Eriksson:2009fj} to make an intermediate step towards FLAIR, providing higher intensities of stopped antiprotons and allowing to gain experience with accelerator and experimental methods in the ultra-low energy regime.



\begin{thebibliography}{10}

\bibitem{Eriksson-LEAP03}
P.~Belochitskii, T.~Eriksson, S.~Maury,
\newblock {\em Nucl. Instr. Meth. Phys. Res. B}
\newblock {\bf  214}, 176 (2004).

\bibitem{Eriksson:2009fj}
Tommy Eriksson,
\newblock {\em Hyperfine Interactions}
\newblock {\bf  194}, 123 (2009).

\bibitem{ATRAP:09}
{G. Gabrielse {\it et al.}}
\newblock {ATRAP} collaboration
\newblock
  http://hussle.harvard.edu/~gabrielse/gabrielse/overviews/Antihydrogen/Antihy%
drogen.html.

\bibitem{ALPHA:09}
{J. Hengst et al.}
\newblock {The ALPHA collaboration}
\newblock http://alpha.web.cern.ch/alpha/.

\bibitem{Niering:00}
M~Niering, R~Holzwarth, J~Reichert, P~Pokasov, Th~Udem, M~Weitz, T.~W.
  H{\"a}nsch, P~Lemonde, G~Santarelli, M~Abgrall, P~Laurent, C~Salomon,
  A~Clairon,
\newblock {\em Physical Review Letters}
\newblock {\bf  84}, 5496 (2000).

\bibitem{ASACUSA}
R.~S. Hayano {ASACUSA} collaboration, http://cern.ch/asacusa/.

\bibitem{Karshenboim:02}
S.~G. Karshenboim,
\newblock In {\em Precision Physics of Simple Atomic Systems},  Eds S.~G.
  Karshenboim, V.~B. Smirnov,
\newblock Springer, Berlin, Heidelberg 2003,
\newblock p. 142,
\newblock {\em hep-ph/0305205}.

\bibitem{HbarLOI}
E.~Widmann, J.~Eades, R.~S. Hayano, T.~Ishikawa, W.~Pirkl, M.~Hori,
  Y.~Yamazaki, A.~Mohri, T.~Yamazaki, D.~Horv{\'a}th, B.~Juh{\'a}sz, E.~Takacs
\newblock Measurement of the antihydrogen hyperfine structure
\newblock Letter of Intent CERN-SPSC-2003-009, CERN, Geneva, Switzerland, 2003.

\bibitem{Iwasaki:91}
M.~Iwasaki, S.~N. Nakamura, K.~Shigaki, Y.~Shimizu, H.~Tamura, T.~Ishikawa,
  R.~S. Hayano, E.~Takada, E.~Widmann, H.~Outa, M.~Aoki, P.~Kitching,
  T.~Yamazaki,
\newblock {\em Phys. Rev. Lett.}
\newblock {\bf  67}, 1246 (1991).

\bibitem{Yamazaki:02}
T.~Yamazaki, N.~Morita, R.~S. Hayano, E.~Widmann, J.~Eades,
\newblock {\em Phys. Rep.}
\newblock {\bf  366}, 183 (2002).

\bibitem{Hayano:2007}
R.S. Hayano, M.~Hori, D.~Horv{\'a}th, E.~Widmann,
\newblock {\em Reports on Progress in Physics}
\newblock {\bf  70}, 1995 (2007).

\bibitem{Hori:01}
M.~Hori, J.~Eades, E.~Widmann, H.~Yamaguchi, J.~Sakaguchi, T.~Ishikawa, R.~S.
  Hayano, H.~A. Torii, B.~Juh\'asz, D.~Horv\'ath, T.~Yamazaki,
\newblock {\em Phys. Rev. Lett.}
\newblock {\bf  87}, 093401 (2001).

\bibitem{Hori:03}
M.~Hori, J.~Eades, R.~S. Hayano, T.~Ishikawa, W.~Pirkl, E.~Widmann,
  H.~Yamaguchi, H.~A. Torii, B.~Juh{\'a}sz, D.~Horv{\'a}th, T.~Yamazaki,
\newblock {\em ibid.}
\newblock {\bf  91}, 123401 (2003).

\bibitem{Hori:2006}
M.~Hori, A.~Dax, J.~Eades, K.~Gomikawa, RS~Hayano, N.~Ono, W.~Pirkl,
  E.~Widmann, HA~Torii, B.~Juh{\'a}sz, et~al.,
\newblock {\em Physical Review Letters}
\newblock {\bf  96}, 243401 (2006).

\bibitem{RFQD-01}
A.~M. Lombardi, W.~Pirkl, Y.~Bylinsky,
\newblock In {\em Proceedings of the 2001 Particle Physics Accelerator
  confenrece},  Eds P.~Lucasa, eds. S.~Webber,
\newblock IEEE, Piscataway, NJ 2001,
\newblock p. 585.

\bibitem{korobovexa05}
V.~I. Korobov,
\newblock In {\em Proceedings of the International Conference on Exotic Atoms
  and Related Topics (EXA'05), Vienna, 2005},  Eds A.~Hirtl, J.~Marton,
  E.~Widmann, J.~Zmeskal,
\newblock Austrian Academy of Sciences Press, Vienna 2005, p. 391.

\bibitem{Kino:04}
Y.~Kino, M.~Kamimura, H.~Kudo,
\newblock {\em Nucl. Instrum. Methods Phys. Res. B}
\newblock {\bf  412}, 84 (2004).

\bibitem{Widmann:02}
E.~Widmann, J.~Eades, T.~Ishikawa, J.~Sakaguchi, T.~Tasaki, H.~Yamaguchi, R.~S.
  Hayano, M.~Hori, H.A. Torii, B.~Juh{\'a}sz, D.~Horv{\'a}th, T.~Yamazaki,
\newblock {\em Phys. Rev. Lett.}
\newblock {\bf  89}, 243402 (2002).

\bibitem{Bakalov:07}
D.~Bakalov, E.~Widmann,
\newblock {\em Physical Review A}
\newblock {\bf  76}, 12512 (2007).

\bibitem{Korobov:2009dq}
Vladimir~I. Korobov, Zhen-Xiang Zhong,
\newblock {\em Physical Review A (Atomic, Molecular, and Optical Physics)}
\newblock {\bf  80}, 042506 (2009).

\bibitem{Pask:2009lq}
T.~Pask, D.~Barna, A.~Dax, RS~Hayano, M.~Hori, D.~Horv{\'a}th, S.~Friedreich,
  B.~Juh{\'a}sz, O.~Massiczek, N.~Ono, et~al.,
\newblock {\em Physics Letters B}
\newblock {\bf  678}, 55 (2009).

\bibitem{Korobov:01}
V.~I. Korobov, D.~Bakalov,
\newblock {\em J. Phys. B: At. Mol. Opt. Phys}
\newblock {\bf  34}, L519 (2001).

\bibitem{PDG:08}
C.~Amsler, et~al.,
\newblock {\em Physics Letters B}
\newblock {\bf  667}, 1 (2008).

\bibitem{Hu:75}
E.~Hu, Y.~Asano, M.~Y. Chen, S.~C. Cheng, G.~Dugan, L.~Lidofsky, W.~Patton,
  C.~S. Wu, V.~Hughes, D.~Lu,
\newblock {\em Nucl. Phys. A}
\newblock {\bf  254}, 403 (1975).

\bibitem{Roberts:75}
{B. L. Roberts {\it et al.}},
\newblock {\em Phys. Rev. D}
\newblock {\bf  12}, 1232 (1975).

\bibitem{Kreissl:88}
A.~Kreissl, A.~D. Hancock, H.~Koch, {Th}. K{\"o}hler, H.~Poth, U.~Raich,
  D.~Rohmann, A.~Wolf, L.~Tauscher, A.~Nilsson, M.~Suffert, M.~Chardalas,
  S.~Dedoussis, H.~Daniel, T.~{von~Egidy}, F.~J. Hartmann, W.~Kanert,
  H.~Plendl, G.~Schmidt, J.~J. Reidy,
\newblock {\em Z. Phys. C}
\newblock {\bf  37}, 557 (1988).

\bibitem{Baur:96}
G.~Baur, G.~Boero, S.~Brauksiepe, A.~Buzzo, W.~Eyrich, R.~Geyer, D.~Grzonka,
  J.~Hauffe, K.~Kilian, M.~LoVetere, M.~Macr{\`\i}, M.~Moosburger, R.~Nellen,
  W.~Oelert, S.~Passaggio, A.~Pozzo, K.~Rohrich, K.~Sachs, G.~Schepers,
  T.~Sefzick, R.~S. Simon, R.~Stratmann, F.~Stinzing, M.~Wolke,
\newblock {\em Phys. Lett. B}
\newblock {\bf  368}, 251 (1996).

\bibitem{De-Beauvoir:2000ru}
B.~De~Beauvoir, C.~Schwob, O.~Acef, L.~Jozefowski, L.~Hilico, F.~Nez,
  L.~Julien, A.~Clairon, F.~Biraben,
\newblock {\em The European Physical Journal D}
\newblock {\bf  12}, 61 (2000).

\bibitem{ATHENA-Hbar:02}
M.~Amoretti, C.~Amsler, G.~Bonomi, A.~Bouchta, P.~Bowe, C.~Carraro, C.~L.
  Cesar, M.~Charlton, M.~J.~T. Collier, M.~Doser, V.~Filippini, K.~S. Fine,
  A.~Fontana, M.~C. Fujiwara, R.~Funakoshi, P.~Genova, J.~S. Hangst, R.~S.
  Hayano, M.~H. Holzscheiter, L.~V. J{\o}rgensen, V.~Lagomarsino, R.~Landua,
  D.~Lindel{\"o}f, E.~{Lodi Rizzini}, M.~Macr{\`\i}, N.~Madsen, G.~Manuzio,
  M.~Marchesotti, P.~Montagna, H.~Pruys, C.~Regenfus, P.~Riedler, J.~Rochet,
  A.~Rotondi, G.~Rouleau, G.~Testera, A.~Variola, T.~L. Watson, D.~P. van~der
  Werf,
\newblock {\em Nature}
\newblock {\bf  419}, 456 (2002).

\bibitem{ATRAP-Hbar:02a}
G.~Gabrielse, N.~S. Bowden, P.~Oxley, A.~Speck, C.~H. Storry, J.~N. Tan,
  M.~Wessels, D.~Grzonka, W.~Oelert, G.~Schepers, T.~Sefzick, J.~Walz,
  H.~Pittner, T.~W. Hansch, E.~A. Hessels,
\newblock {\em Phys. Rev. Lett.}
\newblock {\bf  89}, 213401 (2002).

\bibitem{Gabrielse:2005sl}
G.~Gabrielse,
\newblock {\em Adv. At. Mol. Opt. Phys}
\newblock {\bf  50}, 155 (2005).

\bibitem{Gabrielse:2007ai}
G.~Gabrielse, P.~Larochelle, D.~Le~Sage, B.~Levitt, WS~Kolthammer,
  I.~Kuljanishvili, R.~McConnell, J.~Wrubel, FM~Esser, H.~Gl{\"u}ckler, et~al.,
\newblock {\em Physical Review Letters}
\newblock {\bf  98}, 113002 (2007).

\bibitem{Gabrielse:2008nx}
G.~Gabrielse, P.~Larochelle, D.~Le~Sage, B.~Levitt, WS~Kolthammer,
  R.~McConnell, P.~Richerme, J.~Wrubel, A.~Speck, MC~George, et~al.,
\newblock {\em ibid.}
\newblock {\bf  100}, 113001 (2008).

\bibitem{Andresen:2008wh}
GB~Andresen, W.~Bertsche, PD~Bowe, CC~Bray, E.~Butler, CL~Cesar, S.~Chapman,
  M.~Charlton, J.~Fajans, MC~Fujiwara, et~al.,
\newblock {\em ibid.}
\newblock {\bf  100}, 203401 (2008).

\bibitem{AEGIS}
The {AEgIS} collaboration
\newblock http://aegis.web.cern.ch/aegis/home.html.

\bibitem{Juhasz:ve}
B.~Juh{\'a}sz, E.~Widmann,
\newblock {\em Hyperfine Interactions}
\newblock {\bf  193}, 305 (2009).

\bibitem{MohriEP}
A.~Mohri, Y.~Yamazaki,
\newblock {\em Europhys. Lett.}
\newblock {\bf  63}, 207 (2003).

\bibitem{FLAIR}
{FLAIR} collaboration http://www.oeaw.ac.at/smi/flair/.

\bibitem{FAIR}
{FAIR facility}
\newblock http://www.gsi.de/fair/.

\bibitem{FLAIR-LOI}
{FLAIR} -- a facility for low-energy antiproton and ion research
\newblock Letter of intent, Feb. 2004 available from
  http://www.oeaw.ac.at/smi/flair/.

\bibitem{Danared:2009bz}
H{\aa}kan Danared, Anders K{\"a}llberg, Ansgar Simonsson,
\newblock {\em Hyperfine Interactions}
\newblock {\bf  194}, 129 (2009).

\bibitem{Welsch:2009nk}
C.~Welsch, J.~Harasimowicz, K.~K{\"u}hnel, A.~Papash, M.~Putignano, P.~Schmid,
  J.~Ullrich,
\newblock {\em ibid.}
\newblock {\bf  194}, 137 (2009).

\bibitem{Widmann:2005ys}
E.~Widmann,
\newblock {\em Physica Scripta}
\newblock {\bf  72}, C51 (2005).

\bibitem{grav:2003}
Jochen Walz, Theodor~W. H{\"{a}}nsch,
\newblock {\em General Relativity and Gravitation}
\newblock {\bf  36}, 561 (2004).

\bibitem{Zmeskal:2009ix}
J.~Zmeskal, P.~B{\"u}hler, M.~Cargnelli, T.~Ishiwatari, P.~Kienle, J.~Marton,
  K.~Suzuki, E.~Widmann,
\newblock {\em Hyperfine Interactions}
\newblock {\bf  194}, 249 (2009).

\bibitem{Fahimian:2009zt}
Benjamin Fahimian, John DeMarco, Roy Keyes, Niels Bassler, Keisuke Iwamoto,
  Maria Zankl, Michael Holzscheiter,
\newblock {\em ibid.}
\newblock {\bf  194}, 313 (2009).

\end{thebibliography}
\newcommand{\SortNoop}[1]{}

\end{document}